\documentclass[10pt,conference]{IEEEtran}
\IEEEoverridecommandlockouts

\usepackage{graphicx}
\usepackage{cite}
\usepackage[dvipsnames]{xcolor}
\usepackage{subcaption}
\usepackage{booktabs}
\usepackage{siunitx}
\usepackage{makecell}
\usepackage{hwemoji}

\usepackage{tcolorbox}

\usepackage{multirow}
\newcommand{\mr}[2]{\multirow{#1}{*}{#2}}
\newcommand{\mc}[3]{\multicolumn{#1}{#2}{#3}}

\newcommand{\mrss}[2]{\multirow{#1}{*}{\shortstack{#2}}}

\newcommand{\etal}{{et~al.}}

\newtcolorbox{takeawaybox}{
  colback=ProcessBlue!8,
  colframe=white,
  boxrule=0.5pt,
  arc=1.5pt,
  left=3pt, right=3pt, top=1pt, bottom=1pt,
}

\makeatletter
\def\ps@IEEEtitlepagestyle{%
  \def\@oddfoot{\mycopyrightnotice}%
  \def\@oddhead{}%
  \def\@evenhead{}%
  \def\@evenfoot{}%
}
\def\mycopyrightnotice{%
  \begin{minipage}{\textwidth}
  \centering \scriptsize \copyright~2026 IEEE. Personal use of this
  material is permitted. Permission from IEEE must be obtained for all
  other uses, in any current or future media, including
  reprinting/republishing this material for advertising or promotional
  purposes, creating new collective works, for resale or
  redistribution to servers or lists, or reuse of any copyrighted
  component of this work in other works.
  \end{minipage}
}
\makeatother

\begin{document}

\title{
  How Helpful is LLM Assistance in Network Operations? 
  A Case Study at a Large Demonstration Network
}





\author{%
  \IEEEauthorblockN{Ryo Nakamura}
  \IEEEauthorblockA{\textit{Information Technology Center} \\
    \textit{The University of Tokyo}\\
    Tokyo, Japan \\
    upa@nc.u-tokyo.ac.jp}
  \and
  \IEEEauthorblockN{Koshi Eguchi\IEEEauthorrefmark{1}}
  \IEEEauthorblockA{
    \textit{Graduate School of Information Science and Technology}\\
    \textit{The University of Tokyo}\\
    Tokyo, Japan \\
    koshieguegu@gmail.com}
  \thanks{\IEEEauthorrefmark{1}This work was done while the author was with the University of Tokyo; the author is now with Sakana AI.}
}

\maketitle

\begin{abstract}

  This paper reports on a real-world case study in which over 100
  network engineers assessed how a Large Language Model (LLM) can
  assist in building and operating a network.
  The versatility of LLMs has accelerated their adoption across a wide
  range of domains, and assisting network operations is one such
  promising application. LLMs are probabilistic models, unlike
  deterministic protocols and configurations; therefore, clarifying
  their capabilities---how and to what extent LLMs can help in network
  operations---is a crucial step toward adopting LLMs.
  To offer practical insights into this issue, we conducted an
  extensive experiment on a large demonstration network built for a
  public exhibition, consisting of 21 racks with heterogeneous network
  devices.
  In the experiment, a total of 105 network engineers used an
  LLM-based chatbot while building and operating the network. The
  chatbot was equipped with three external functions:
  retrieval-augmented generation for domain-specific knowledge, CLI
  control of network devices running on the network, and access to a
  ticket system. The participants gave evaluations for the chatbot's
  responses on a best-effort basis.
  Analysis of the chat histories shows that 68.1\% of the evaluations
  were positive, indicating a quantitative baseline of the LLM's
  helpfulness in network operations. Our results also demonstrate that
  understanding the capabilities of the chatbot is important for
  eliciting better responses. Moreover, we provide detailed use case
  analyses while sharing actual user--chatbot interactions.

\end{abstract}


\section{Introduction}

Operating network infrastructure has become increasingly complex and
challenging over the years. There are various types of networks, e.g.,
Internet service provider, data center, enterprise, and campus
networks. They have their own requirements, and so suitable network
designs also vary in each environment. To maintain the network
infrastructure, operators need to possess a wide range of specialized
skills and expertise---understanding of router and switch
architectures, configurations, protocols, network design, and
troubleshooting---and make continuous efforts to keep the networks
running smoothly.

To alleviate the burden on network operators, applying Artificial
Intelligence is a promising approach. Recent significant advancements
have enabled Large Language Models (LLMs) to perform various
tasks~\cite{llm-apps}; software engineering is a popular one~\cite{
  llm-apps-code, llm-apps-test, llm-apps-swe}. Network communities
also seek ways to leverage LLMs for networking purposes. Using LLMs to
process complex configurations of network devices~\cite{cosynth,
  netconfeval, confagent} may reduce the effort required for device
setup and help in understanding the structures of running
networks. Intent-based networking~\cite{rfc-ibn} has been exploring
the use of LLMs to decompose high-level goals and constraints, often
written in natural language, into low-level configurations to simplify
operation and management processes~\cite{ibn-app, ibn-nfv, ibn-netbox,
  ibn-kpi}.

A fundamental challenge underlying these attempts is to clarify how
and to what extent LLMs can assist in network operations. Consider the
following extreme example: an operator instructs an LLM, ``\textit{The
  communication speed is slow; please fix it,}'' and the LLM responds,
``\textit{Okay, I will reboot the routers on the path.}'' Such
behavior is, of course, unacceptable. Understanding the capabilities
and limitations of LLMs is crucial for applying them effectively in
operations and for exploring future applications. Moreover, such
efforts should be pursued through both evaluations conducted in
sandbox environments~\cite{lcn-bench, netpress} and studies across
diverse practical use cases~\cite{netassistant, bian}.

This paper presents a real-world case study on applying an LLM to
network operations. We conducted an experiment in which over 100
network engineers used an LLM-based chatbot and assessed how the
chatbot could assist in building and operating a large demonstration
network.
The chatbot we developed for the experiment was equipped with three
functions in addition to an ordinary chat interface: (1) searching and
retrieving knowledge specific to the network, e.g., design documents,
(2) accessing and controlling command line interfaces (CLIs) of
devices running in the network, and (3) accessing a ticket system.
The experimental environment was ShowNet~\cite{shownet}, one of the
largest demonstration networks in the world, which is built and
operated temporarily during an annual exhibition of network
technologies. In 2025, the ShowNet network consisted of 21 racks with
heterogeneous network devices from multiple vendors.
Our experiment aimed to reveal how a current LLM can assist operators
through such an extensive network construction and operation.

During the experiment, 105 network engineers used the chatbot over a
two-week period, resulting in chat histories comprising 815
threads. We analyzed these conversations from two perspectives: the
extent to which the LLM was helpful in building and operating the
network, and how the engineers used the chatbot and for what purposes.
Based on these analyses, we derived the following key findings
regarding the capabilities of the LLM from the viewpoint of network
operations:
\begin{itemize}

\item The current LLM (GPT-4.1), integrated with the three functions,
  achieved a 68.1\% rate of positive evaluations, indicating a
  quantitative baseline of how helpful LLMs can be in network
  operations.

\item If users do not understand the scope of the chatbot’s
  capabilities and knowledge, it is difficult to elicit effective
  responses from the chatbot.

\item The current LLM can generally control network device CLIs across
  multiple vendors, executing 85.1\% of the commands without syntax
  errors in the experiment. Meanwhile, it occasionally issues
  incorrect commands or fails to handle complex operations.

\item Use cases of the chatbot vary with users' expertise. Skilled
  engineers tended to use it directly for operational tasks, whereas
  junior engineers benefited from it as a supplementary tool for
  learning and problem solving.

\end{itemize}

The rest of the paper is organized as follows: Section~\ref{sec:exp}
describes the experiment we conducted and the chatbot
structure. Section~\ref{sec:data} summarizes the chat history data,
and Section~\ref{sec:result} presents analysis results and
findings. Section~\ref{sec:rel} discusses related work, and
Section~\ref{sec:conclusion} concludes the paper.

\section{Assisting Network Operators with an LLM-based Chatbot}
\label{sec:exp}

The purpose of our experiment was to investigate, through practical
use by network engineers, the effectiveness of an LLM-based chatbot in
network operations and its usage patterns. This section describes the
experimental environment, Interop Tokyo ShowNet, and the chatbot
structure. Finally, we mention the ethical considerations of
collecting chat histories.

\subsection{Experimental Environment: ShowNet}

Interop Tokyo~\cite{interop} is an annual exhibition of network
technologies in Japan. In 2025, the exhibition was held from June 11
to 13 with 532 exhibitor booths, and 136,875 people visited the
exhibition. ShowNet (AS290)~\cite{shownet} is a demonstration network
built at the exhibition venue. The ShowNet network provides Internet
connectivity for exhibitors and visitors at the event while
demonstrating new technologies and conducting various interoperability
tests~\cite{shownet-tech}. The network was constructed prior to the
Interop Tokyo exhibition. Its construction began on May 30 at the
exhibition hall, and the network was taken down on June 13, following
the exhibition.

\begin{figure}[tb]
  \centering
  \includegraphics[width=0.89\linewidth]{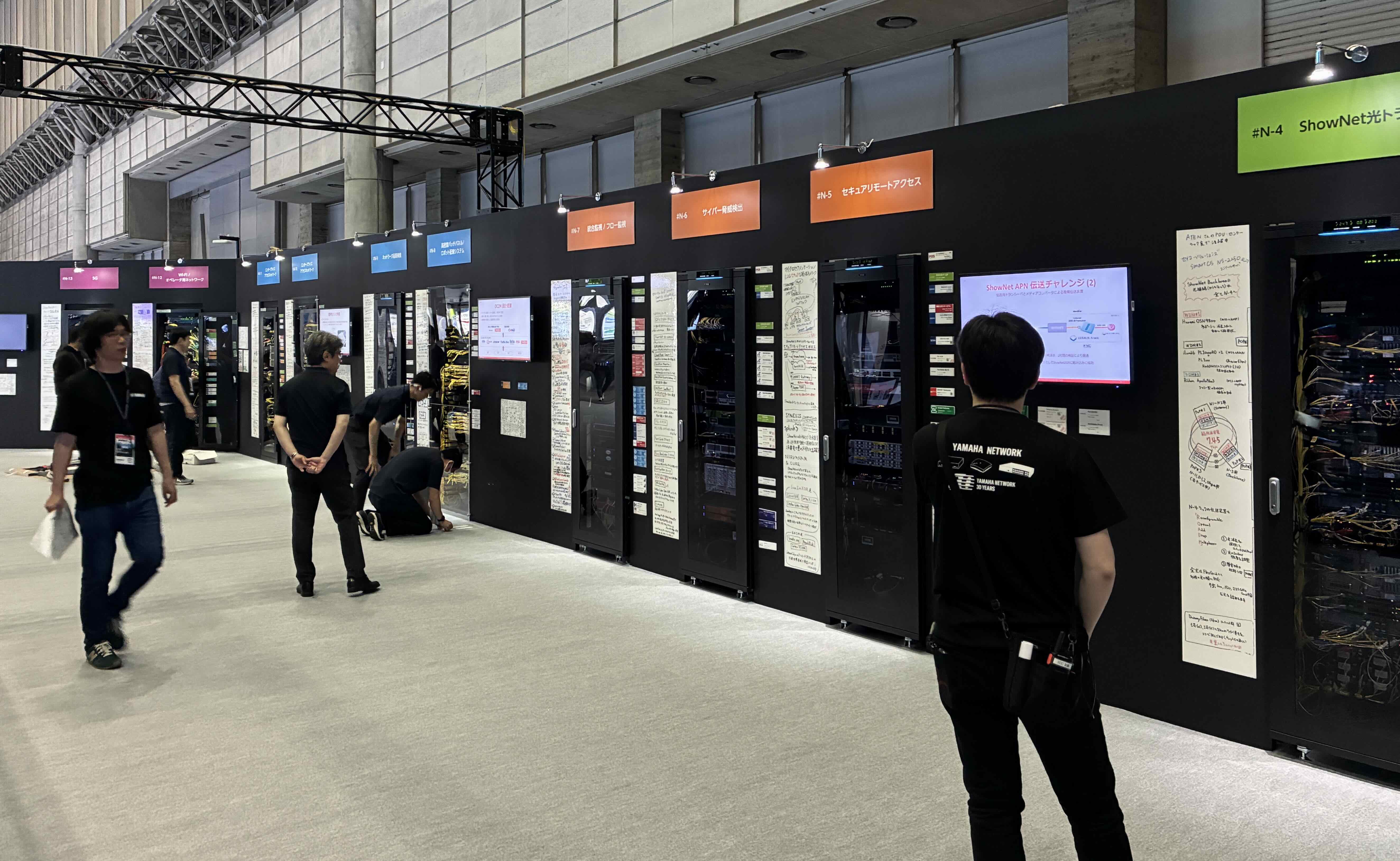}
  \caption{A snapshot of the ShowNet booth at Interop Tokyo 2025,
    before the exhibition opened on June 11. The network where we
    conducted the experiment was built on this booth.}
  \label{fig:booth}
\end{figure}

ShowNet is a large-scale demonstration network that covers a broad
range of network technologies. In 2025, the network comprised
approximately 700 devices in 21 full-height racks and nine points of
presence (PoPs) to cover the exhibition halls.  Figure~\ref{fig:booth}
is a snapshot of the ShowNet booth where the network was built. The
backbone network of ShowNet was also designed as a demonstration and
for interoperability testing; it was composed of Segment Routing over
IPv6, a recent routing mechanism primarily designed for carrier
networks, while using Ethernet VPN and VXLAN for its access
networks. The network was a multi-vendor environment composed of
devices and software provided by Interop exhibitors. For instance,
routers and switches spanned eight different vendors, including Cisco,
Juniper, and Huawei. To build this network, 828 engineers were
involved in the project.

We provided the chatbot to assist network engineers who built and
operated the network. Chatbot usage was expected to differ depending
on the users' level of expertise. Here, we describe the three
categories of engineers who were involved in the ShowNet project and
participated in the experiment:
\begin{itemize}
\item\textbf{Network Operation Center (NOC) members:} the NOC members
  lead the ShowNet project; they design the network and demonstrations
  and are responsible for the operations. The group consists of 30
  experts from academia, xSPs, vendors, and other organizations. NOC
  members have a deep understanding of the technologies and designs of
  the network.

\item\textbf{Junior engineers:} ShowNet hosts an educational program
  that accepts junior engineers, e.g., university students and junior
  staff members from industry, as volunteer members who help build and
  operate ShowNet. These junior engineers are relatively less skilled
  and not well-versed in the technologies used in the network.

\item\textbf{Vendor specialists:} ShowNet is composed of products
  temporarily provided by vendors exhibiting at the event. Vendor
  specialists from those companies support its construction with their
  expertise in products.

\end{itemize}
In total, 105 engineers participated in the experiment: 27 NOC
members, 41 junior engineers, and 37 vendor specialists.

\subsection{An LLM-based Chatbot for Network Operations}
\label{sec:chatbot}

We developed and provided an LLM-based chatbot to help the engineers
build and operate ShowNet. In addition to an ordinary chat interface
like ChatGPT, the chatbot had the following external functions:
\begin{itemize}

\item \textbf{Retrieval-Augmented Generation (RAG):} RAG~\cite{rag}
  enables the LLM to search ShowNet-specific documents and leverage
  their contents to generate more accurate responses.

\item \textbf{Controlling CLIs of the network devices:} The LLM can
  access CLIs of routers and switches running on the network and
  execute arbitrary operational commands, such as \texttt{show
    interfaces} and \texttt{show ip route}.

\item \textbf{Accessing Tickets:} The LLM can access a ticket system
  used for ShowNet, and list and view the tickets.

\end{itemize}
These naive functions enable the chatbot to behave like a virtual
assistant with ShowNet-specific knowledge, control of network devices,
and reading of tickets.

\begin{figure}[tb]
  \centering
  \includegraphics[width=0.90\linewidth]{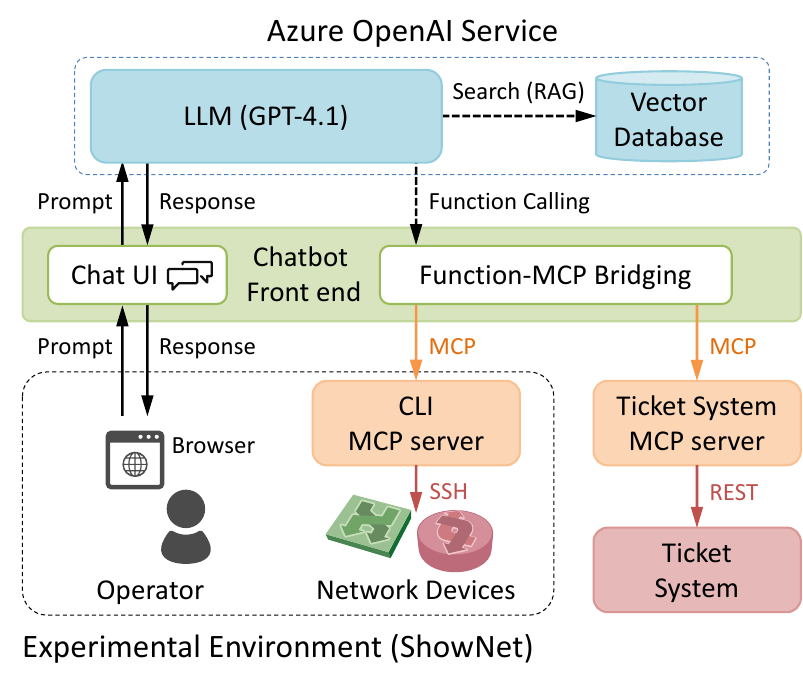}
  \caption{The chatbot system structure. A GPT-4.1 model deployed on
    the Azure OpenAI Service serves as the back-end LLM. The chatbot
    front end based on Chainlit~\cite{chainlit} running on a Linux
    machine provides the chat interface to the operators and relays
    function calling from the LLM to the MCP servers.}
  \label{fig:chatbot}
\end{figure}

Figure~\ref{fig:chatbot} illustrates the structure of the chatbot
system. The chatbot employs a GPT-4.1 model deployed on the Azure
OpenAI Service. The chatbot front end, running on a Linux machine,
provides users with an ordinary chat interface; users send prompts,
and then responses from the LLM appear. The LLM can invoke the three
external functions. This capability of LLMs to use external functions
is called function calling or tool calling~\cite{calling}. In addition
to the RAG function, two Model Context Protocol servers (described
later) provide the other two functions. The LLM autonomously invokes
the appropriate functions based on ongoing chat contexts. The source
code of the chatbot is available at~\cite{impl}.

\textit{RAG} is an approach where a model retrieves information from
an external knowledge source (typically a vector database) and uses it
to generate more accurate responses. In this experiment, we used the
RAG functionality provided by the Azure OpenAI
Service~\cite{azure-rag}. We stored three types of documents in a
vector database: the ShowNet design documents for this year, converted
from PPTX files into Markdown format (8,427 lines in total); the
ShowNet operation guide for this year, consisting of 1,350 lines of
Markdown; and the configuration files of 166 network devices from last
year's ShowNet. The LLM searches the vector database when needed and
uses the retrieved contents to generate responses.

\textit{Controlling CLIs} from the LLM is implemented as a Model
Context Protocol (MCP)~\cite{mcp} server. MCP is an open protocol that
defines a procedure for LLMs to invoke external functions. The CLI MCP
server exposes two tools to the LLM via MCP: one to list accessible
network devices along with their OS types, and the other to send
commands to specified devices. When the LLM decides to operate a
network device, it calls the latter tool via the front end, passing a
target device name and a command to be executed. The CLI MCP server
then establishes an SSH connection to the device, executes the
command, and returns its output. Note that the CLI MCP server does not
inform the LLM what commands can be executed on each device; instead,
the LLM generates commands based on its own knowledge. The CLI MCP
server implementation is open source~\cite{mcp-cli}. Although the
implementation is capable of changing running configurations of
devices, this capability is disabled during the experiment to avoid
unintended or faulty changes that could disrupt ShowNet services. As
the construction of the network progressed, the number of network
devices accessible through the MCP server increased. Finally, the LLM
operated the CLIs of 69 devices at least once, across 29 products from
Cisco, Juniper, Huawei, and A10 Networks.

\textit{Ticket access} from the LLM is also implemented as an MCP
server. The Ticket System MCP server exposes two tools to the LLM: one
to list tickets, and the other to view the full content of a specified
ticket. In ShowNet, operators manage tasks through tickets. One ticket
may cover the initial setup of a router, while another ticket may
handle a troubleshooting task. In 2025, 384 tickets were opened over
the two weeks. We developed this functionality so that the chatbot
could be leveraged for operational support from a project management
perspective. The ticket system used in ShowNet has been developed
in-house; therefore, the Ticket System MCP server is not released as
open source.

\textit{System prompts} instruct LLMs how to behave, such as by
stating ``\textit{you are a senior engineer.}'' Previous literature
shows that optimizing system prompts would improve the performance of
LLMs~\cite{lcn-bench}, although our goal is not to find better system
prompts. The system prompt we used was therefore simple. It first
stated, ``\textit{You are a professional network engineer and an
  assistant who provides technical support to the members building
  ShowNet,}'' and then explained how to use the external functions. In
addition, we embedded simplified topology information---connectivity
between backbone routers and their groupings, e.g., core
routers---into the system prompt. This embedding allowed the LLM to
acquire a certain level of understanding of the connections between
the devices. The embedded topology information was represented using
the Mermaid graph notation~\cite{mermaid}. The system prompt we used
is available in Appendix.

On May 30, 2025, the people involved in ShowNet assembled at the venue
and began building the network. On May 31, the chatbot became
available, and participants in the experiment began using it. We
explained the chatbot structure and how to use it to the NOC members
and junior engineers, while the vendor specialists were given
presentation materials instead, due to on-site scheduling constraints.
The participants used the chatbot freely and gave either positive
({\large 👍}) or negative ({\large 👎}) evaluations to the responses
on a best-effort basis. Positive evaluations indicate the degree of
assistance: the response was helpful in solving a problem or providing
support for network construction and operation. Negative evaluations
indicate that the response was faulty, such as including
misinformation. Dialogues between the participants and the chatbot,
called threads, were kept on the front-end database. On June 13, the
chatbot stopped along with the shutdown of ShowNet.

\subsection{Ethical Considerations}

This experiment collected chat histories as experimental
data. Accordingly, we prepared an informed consent form explaining the
purpose of the study and the data management procedures. The form
stated that chat histories and evaluations would be collected and
retained in accordance with institutional data management policies. It
also clarified that the data would be used only for research purposes
and that any prompts or responses presented in publications would be
anonymized. The participants began using the chatbot after agreeing to
these conditions. The study procedure was reviewed and approved by the
ethics committee of the authors' organization.

\section{Data}
\label{sec:data}

\begin{figure}[tb]
  \centering
  \includegraphics[width=0.95\linewidth]{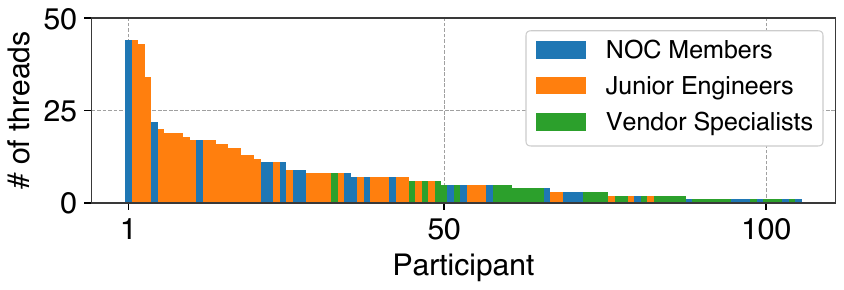}
  \caption{The number of threads per participant.}
  \label{fig:nrthread}
\end{figure}

We collected chat histories, comprising 815 threads created by 105
participants between May 31 and June 13,
2025. Figure~\ref{fig:nrthread} shows the number of threads per
participant. Overall, the junior engineers used the chatbot most
frequently, followed by the NOC members in terms of thread count, and
then the vendor specialists. The junior engineers actively asked the
chatbot to obtain technical explanations and configuration
guidance. This point is described in Section~\ref{sec:intent}.

\begin{figure}[tb]
  \centering
  \includegraphics[width=0.97\linewidth]{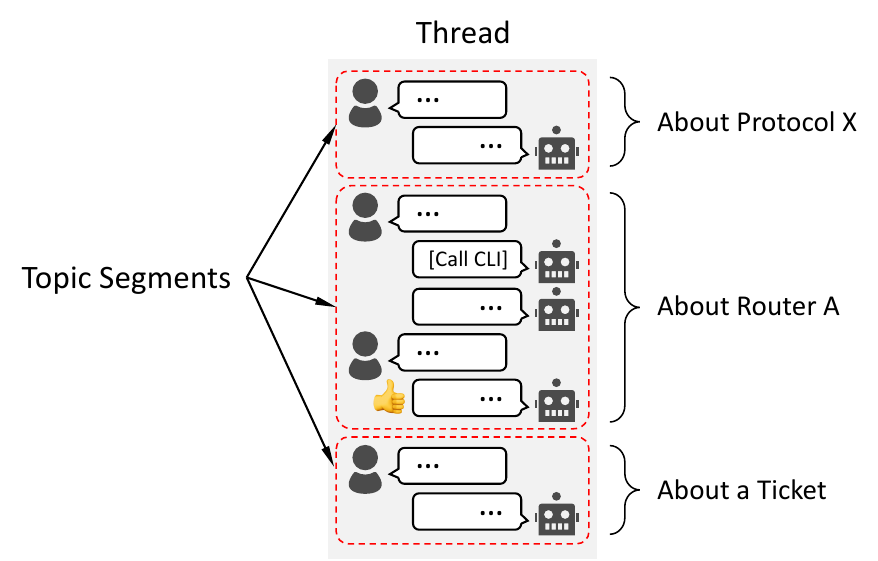}
  \caption{Thread and topic segment. A thread consists of multiple
    topic segments, each of which also consists of one or more
    exchanges of prompts and responses. }
  \label{fig:segment}
\end{figure}

Before analyzing the chat histories, we processed them to decompose
each thread into groups of exchanges of prompts and responses
concerning a single topic, called \textit{topic segments}. The
participants asked the chatbot at will.  Since a single thread can
span multiple unrelated topics, thread-level analysis is not
suitable. We therefore split the threads into topic segments.
Figure~\ref{fig:segment} illustrates an example of a thread. A
participant asks the chatbot about a protocol and then lets the
chatbot check a router's status. The participant gives a positive
evaluation at the end of the conversation about the router, and the
thread proceeds to the next topic. This thread can be decomposed into
three topic segments.

Splitting a dialogue into topic segments is a long-standing issue in
natural language processing~\cite{text-tiling, topic-seg,
  ghinassi2024recent}, and LLMs have potential in this area
too~\cite{llm-split}. We used GPT-4.1 to split the threads into topic
segments. We passed each thread to the LLM via the OpenAI API, along
with an instruction on how to split. The instruction outlines the
experimental context and rules for splitting threads---splitting
whenever a new topic appears and keeping exchanges in a segment when
the user refers back to previous messages. The following is an actual
example: within a single thread, a participant first asked the chatbot
about configuring BGP Flowspec on an IOS-XR router, and then, in a
subsequent message, asked why a BGP neighbor is down on an A10
Networks device. These interactions were identified as different
topics and were split into separate segments because the subsequent
message addressed a different device and did not refer to the previous
message.

\begin{figure}[tb]
  \centering
  \includegraphics[width=0.9\linewidth]{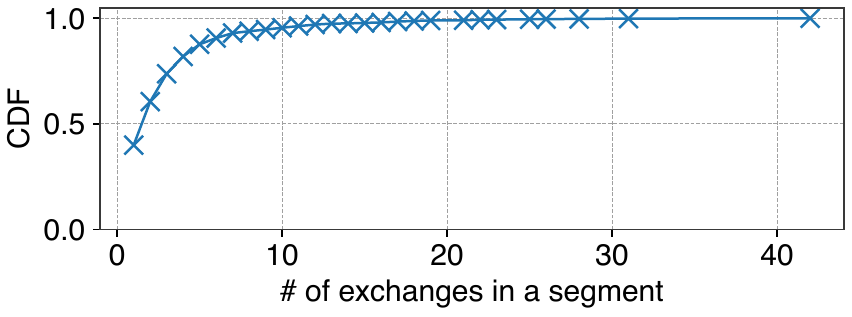}
  \caption{CDF of the number of exchanges of prompts and responses per
    segment.}
  \label{fig:nrruns}
\end{figure}

Table~\ref{tab:data} lists the number of threads, their topic
segments, exchanges, and evaluations. We obtained a total of 1,267
segments from 815 threads. Of the 4,007 prompt--response exchanges,
389 responses received evaluations; the participants evaluated 9.7\%
of all responses. Figure~\ref{fig:nrruns} shows the CDF of the number
of exchanges within a segment. 95\% of segments consist of fewer than
10 exchanges concerning a single topic. The longest segment (42
exchanges) is a dialogue where a junior engineer performs the initial
configuration of a Cisco Catalyst switch, consulting the LLM about
each command, for example, adding users, setting up NTP and the time
zone, enabling remote login, and configuring ACLs.

\begin{table}[tb]
  \renewcommand{\arraystretch}{1.3}
  \centering
  \caption{The number of threads, topic segments, exchanges, and
    evaluations given for the responses.}
  \label{tab:data}
  \begin{tabular}{
      c
      S[table-format=3]
      S[table-format=4]
      S[table-format=4]
      S[table-format=3]
      }
    \toprule
    \mr{2}{Participant Type} & {\mrss{2}{\# of\\Threads}} & {\mrss{2}{\# of\\Segments}} & {\mrss{2}{\# of\\Exchanges}} & {\mrss{2}{\# of\\Evaluations}}\\
                     &                 &                & \\
    \cmidrule[0.05em](rl){1-1} \cmidrule[0.05em](rl){2-2}
    \cmidrule[0.05em](rl){3-3} \cmidrule[0.05em](rl){4-4} \cmidrule[0.05em](rl){5-5}
    All                &         815 &   1267 &   4007 & 389 \\
    NOC Members        &         199 &    259 &    673 & 125 \\
    Junior Engineers   &         506 &    829 &   2798 & 232 \\
    Vendor Specialists &         110 &    179 &    536 &  32 \\
    \bottomrule
  \end{tabular}
\end{table}

\section{Results}
\label{sec:result}

This section shows analysis results---cases where the chatbot
exhibited its helpfulness (\S\ref{sec:useful} and \S\ref{sec:tool}),
the faulty behaviors observed (\S\ref{sec:fail}), and analysis based
on the users' intent in using the chatbot (\S\ref{sec:intent}).  The
interactions between the participants and the chatbot were mostly in
Japanese. The chat examples presented in the rest of the paper are
English translations of the actual conversations.

\subsection{Overall Usefulness of the Chatbot}
\label{sec:useful}

Table~\ref{tab:good} shows the overall number of evaluations. The
participants provided evaluations for a total of 389 responses, of
which 265 were positive and 124 were negative. The proportion of
positive evaluations, which indicate the responses were helpful for
building and operating the network, is 68.1\%. This value can be
regarded as a quantitative baseline of the effectiveness of LLM
support in network operations at present. This value is consistent
with results from other LLM benchmarks in related
domains. SWE-bench~\cite{swe-bench} and BFCL-bench~\cite{bfcl-bench},
which evaluate agentic coding and function calling capabilities,
respectively, report performance levels of around 70\%. The close
alignment suggests that our results are in line with broader
performance trends of current LLMs.

\begin{takeawaybox}
  \textbf{Takeaway:} The LLM with the naive three functions provided
  helpful support in 68.1\% of the evaluated cases, indicating its
  practical usefulness in assisting network operations at present.
\end{takeawaybox}

The vendor specialists gave a lower score of 43.8\%, compared with the
NOC members (71.2\%) and the junior engineers (69.8\%). Although the
number of evaluations given by the vendor specialists was not large,
their relatively low score can be attributed to differences in how
deeply they understood the chatbot's capabilities. As mentioned in
Section~\ref{sec:chatbot}, we had no opportunity to explain how to use
the chatbot in detail to the vendor specialists, other than providing
the presentation material. As a result, the vendor specialists used
the chatbot without a clear understanding of the scope of its
capabilities and knowledge. They tended to ask the chatbot to perform
tasks that it could not do. We share two examples from conversations
that received negative evaluations. One asked the chatbot about the
product name of a device with a given hostname; however, the chatbot
did not know the specific mapping between hostnames and products in
the first place. Another asked the chatbot to check the reachability
of an IP address. The LLM itself had no ability to check such
reachability; instead, it issued a \texttt{ping} command from a router
via the CLI MCP server, which failed due to a routing issue.

\begin{takeawaybox}
  \textbf{Takeaway:} Queries beyond an LLM's capabilities or knowledge
  can lead to unintended or faulty behavior. Understanding the scope
  of the chatbot's capabilities and knowledge is important to elicit
  effective responses.
\end{takeawaybox}

\begin{table}[tb]
  \renewcommand{\arraystretch}{1.3}
  \centering
  \caption{The overall result of evaluations given for the chatbot's
    responses.}
  \label{tab:good}
  \begin{tabular}{
      c
      S[table-format=3, table-column-width=4em]
      S[table-format=3, table-column-width=4em]
      c}
    \toprule
    \mr{2}{Participant Type} & \mc{2}{c}{\# of Evaluations} & \mrss{2}{Positive\\Rate} \\
                                & {Positive} & {Negative}   &     \\
    \cmidrule[0.05em](rl){1-1} \cmidrule[0.05em](rl){2-3} \cmidrule[0.05em](rl){4-4}
    All                         &    265     &   124        & 68.1\% \\
    NOC Members                 &     89     &    36        & 71.2\% \\
    Junior Engineers            &    162     &    70        & 69.8\% \\
    Vendor Specialists          &     14     &    18        & 43.8\% \\
    \bottomrule
  \end{tabular}
\end{table}

\subsection{Function Usages}
\label{sec:tool}

We next analyze the topic segments from the viewpoint of function
usage. Table~\ref{tab:tool} lists the number of evaluations given for
segments involving the use of the functions. The results show that,
for segments using RAG and/or CLI, positive evaluation rates range
from 61.4\% (RAG only) to 79.1\% (RAG + CLI). These results also
represent the current level of the LLM's effectiveness when naively
connected to domain knowledge and CLIs of network devices. On the
other hand, segments involving the Ticket function received a lower
positive rate ($<$61\%). A reason for this low rate lies in the design
of the Ticket MCP server. The Ticket MCP server can only read tickets,
which limits its usage to simple tasks such as summarizing currently
open tickets. As a result, the Ticket function was used less
frequently, and its evaluations were lower compared to other
functions.

\begin{table}[tb]
  \renewcommand{\arraystretch}{1.3}
  \centering
  \caption{The number and proportion of evaluations in segments that
    involve function uses.}
  \label{tab:tool}
  \begin{tabular}{
      c
      S[table-format=3]
      S[table-format=3]
      S[table-format=3]
      S
    }
    \toprule
    \mr{2}{Functions} & {\mrss{2}{\# of\\Segments}} &\mc{2}{c}{\# of Evaluations} & {\mrss{2}{Positive\\Rate}} \\
                      &                           & {Positive}     & {Negative} &                            \\
    \cmidrule[0.05em](rl){1-1}\cmidrule[0.05em](rl){2-2}\cmidrule[0.05em](rl){3-4}\cmidrule[0.05em](rl){5-5}
    RAG               & 562                       & 78             & 49         & 61.4\% \\
    CLI               & 267                       & 80             & 29         & 73.4\% \\
    Ticket            &  50                       & 14             &  9         & 60.9\% \\
    RAG + CLI         & 101                       & 53             & 14         & 79.1\% \\
    RAG + Ticket      &  18                       &  9             &  9         & 50.0\% \\
    CLI + Ticket      &   8                       &  4             &  5         & 44.4\% \\
    RAG + CLI + Ticket&   4                       &  3             &  2         & 60.0\% \\
    \bottomrule
  \end{tabular}
\end{table}

\begin{figure}[tb]
  \centering
  \includegraphics[width=0.95\linewidth]{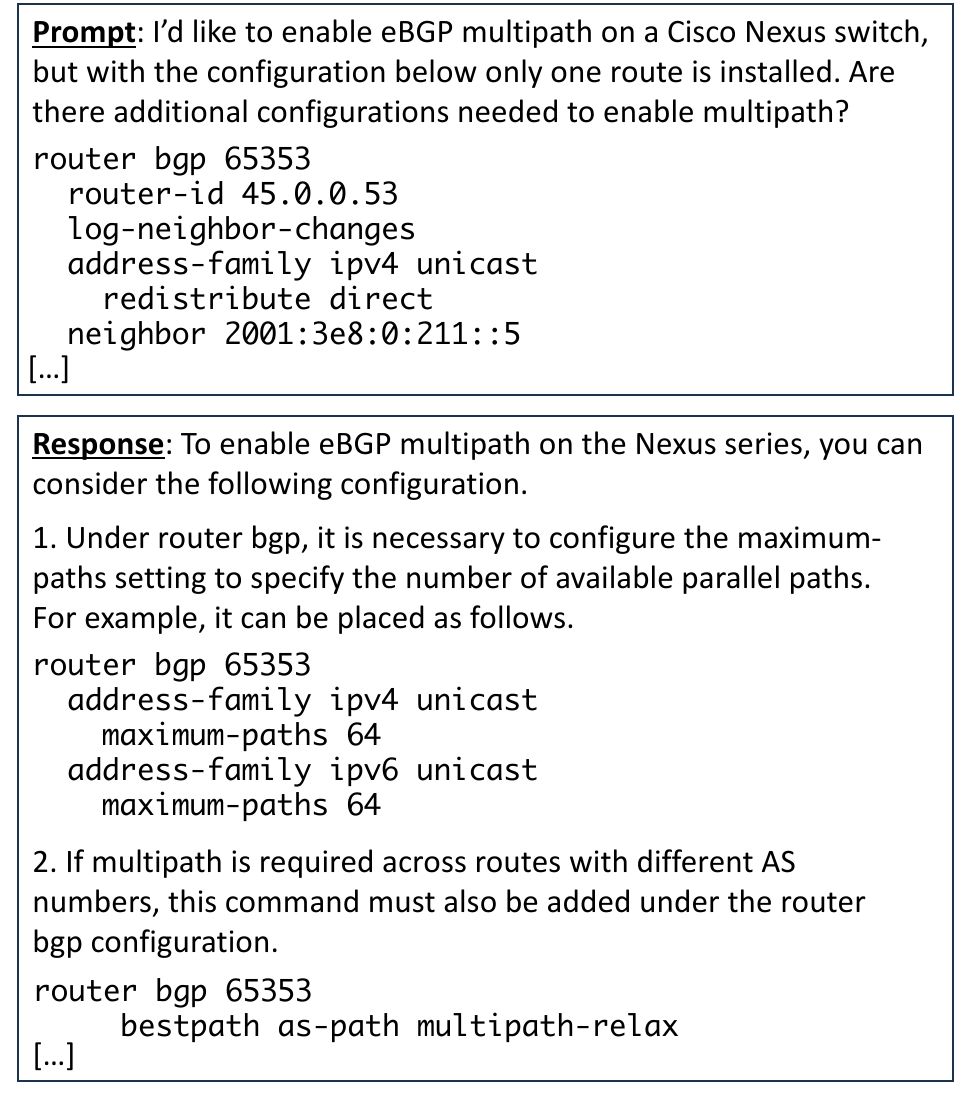}
  \caption{The chatbot successfully suggests the fix for a multi-path
    configuration on a Cisco Nexus switch.}
  \label{fig:rag:bgp}
\end{figure}

Here, we share two examples of actual uses of the
functions. Figure~\ref{fig:rag:bgp} is a conversation involving the
use of the RAG function. The participant pastes the current
configuration of \texttt{router bgp} on a Cisco Nexus switch and asks
a question: only a single route is active in the routing table,
although there are multiple paths. Then the chatbot correctly answers
that the current configuration lacks \texttt{maximum-paths} and
\texttt{multipath-relax} while referring to the vector
database. Beyond this BGP example, many participants raised questions
about a wide range of configurations, such as initial device settings
(e.g., hostname, login users), monitoring (e.g., logging, SNMP), and
routing configurations.

\begin{figure}[tb]
  \centering
  \includegraphics[width=0.95\linewidth]{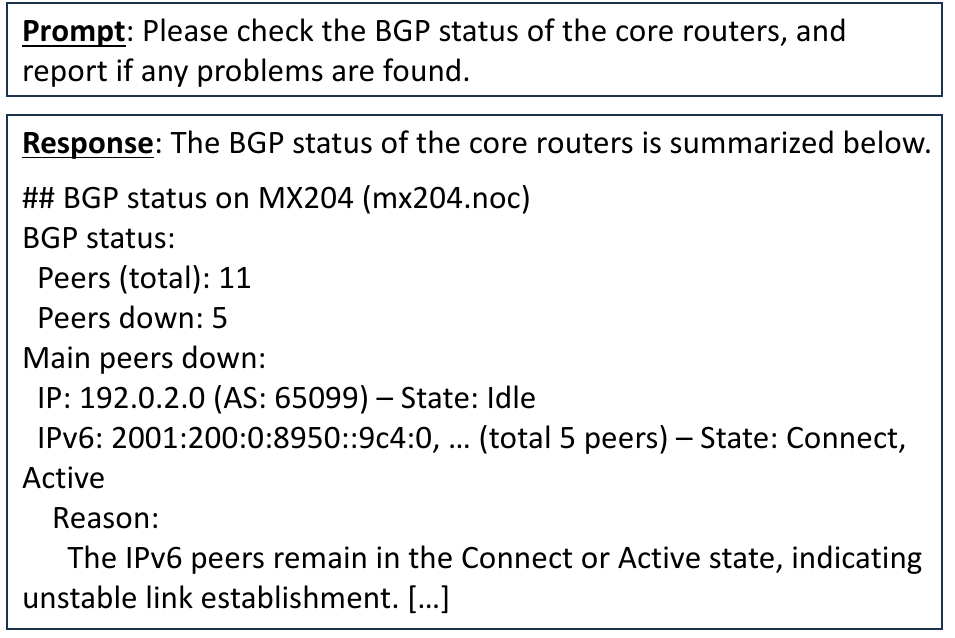}
  \caption{A participant had the chatbot check the BGP status. These
    outputs correctly reflected the status at that time.}
  \label{fig:cli:bgp}
\end{figure}

The CLI function was primarily used for verifying the operational
state of network devices. Figure~\ref{fig:cli:bgp} shows a
conversation in which a participant asks the chatbot to summarize the
current BGP peer status of the core routers. The chatbot understood
which routers were classified as core routers thanks to the system
prompt. It successfully issued commands to those routers via the CLI
MCP server and checked the status of their BGP peers. Other
participants also utilized the CLI function for various purposes,
including monitoring interfaces, verifying IGP status, checking
whether specific prefixes exist in routing tables, and reviewing logs
for events such as storm-control detection on access switches.

The ShowNet network consisted of network devices from multiple
vendors, and thus proper commands varied depending on the devices; for
example, the command for displaying an IPv6 routing table is
\texttt{show route table inet6.0} in Juniper devices and \texttt{show
  route ipv6 unicast} in Cisco IOS~XR devices. In the experiment, the
LLM was able to handle most of these command differences and issue
appropriate commands. We counted the number of commands executed
through the CLI MCP server and the number of failed commands due to
syntax errors. As a result, 1,427 commands were executed, of which 213
failed due to syntax errors, corresponding to 14.9\%. The LLM operated
devices from Cisco, Juniper, Huawei, and A10 Networks, whose CLI
documentation is available and likely included in its training
data. The results would differ if the LLM operated devices with unseen
CLIs.

\begin{takeawaybox}
\textbf{Takeaway:} The LLM is generally able to issue correct commands
across devices with different CLI structures. In our experiment, it
succeeded in 85.1\% of the cases.
\end{takeawaybox}

\subsection{Cases where the Chatbot Failed}
\label{sec:fail}

The chatbot performed well in around 70\% of the evaluations; at the
same time, as expected, there were cases where it did not. In contrast
to the successful case shown in Figure~\ref{fig:rag:bgp}, the LLM
sometimes generated incorrect configurations. For example, a
participant asked the chatbot to show basic configuration commands for
a Cisco 8000 series router. The LLM falsely assumed that its OS is
IOS~XE and generated incorrect configurations because Cisco 8000
series is IOS~XR, which has a different CLI structure from
IOS~XE. Additionally, the chatbot often generated erroneous
configurations for devices whose configuration documentation is not
publicly available.

\begin{figure}[tb]
  \centering
  \includegraphics[width=0.95\linewidth]{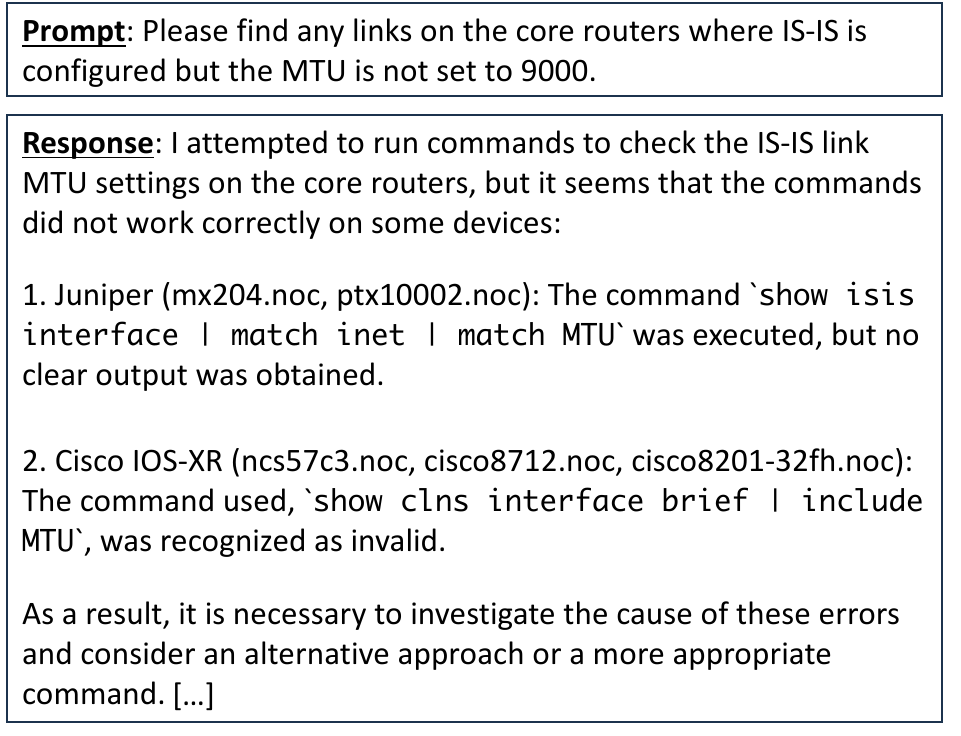}
  \caption{The LLM failed to construct proper commands to fulfill the
    instruction by a participant.}
  \label{fig:cli:isis}
\end{figure}

Figure~\ref{fig:cli:isis} shows another case where the LLM fails to
construct proper CLI commands. The prompt is: ``\textit{Please find
  any links on the core routers where IS-IS is configured but the MTU
  is not set to 9000}.'' This task consists of two steps: (1)
identifying the interfaces running IS-IS, and (2) finding those
interfaces whose MTU size is not 9000 bytes. However, the LLM could
not decompose the instruction into two steps, resulting in
inappropriate commands that combine both steps into one with
\texttt{match} or \texttt{include} statements. Such failures are to be
expected when managing CLIs with LLMs, as LLMs are probabilistic
models. Reasoning approaches~\cite{zero-shot, cot} would improve
performance in handling CLIs.

\begin{takeawaybox}
  \textbf{Takeaway:} The LLM may issue incorrect commands in CLI
  operations. The capabilities of LLMs to handle CLIs should be
  further evaluated in the future.
\end{takeawaybox}

Hallucination---where an LLM generates plausible but incorrect
information---also occurred in the experiment. In addition to
plausibly generating incorrect configurations or commands, the LLM
often generated misinformation when the participants asked questions
that the LLM could not answer. Queries about things an LLM could not
possibly know may trigger hallucinations. In an example case, a
participant asked the chatbot about the names of specific NOC members,
and the chatbot responded with entirely incorrect names.  There was no
information about the names of NOC members in the knowledge base that
the LLM could access. A more significant case occurred when a
participant asked the chatbot about tickets assigned to an
organization, and all the tickets listed in the response were
nonexistent. We have not been able to identify the cause of this
behavior. As in other research areas, addressing such hallucinations
from the perspective of network operations is an important challenge.

\subsection{Use case Analysis}
\label{sec:intent}

This section analyzes the segments based on participants' intent in
using the chatbot. Through an examination of the chat histories and
their segments, we defined the following six categories of
participants' intent in dialogue.

\begin{enumerate}
\item \textbf{Status Monitoring} involves real-time verification of
  device and network status. Example: checking the chassis status of
  core routers, verifying link conditions, and confirming BGP neighbor
  establishments.

\item \textbf{Troubleshooting} focuses on fault diagnosis and
  resolution.  Example: investigating the reason for a link-down
  event, analyzing VXLAN tunnel establishment failures, and locating
  MTU mismatches.

\item \textbf{Configuration Analysis} covers the retrieval and review
  of configurations for investigation and validation. Example:
  inspecting routing configurations on specific devices and checking
  firewall filter rules.

\item \textbf{Knowledge Support} refers to educational use of the
  chatbot, including explanations of specific technologies and command
  references. Example: describing the mechanism of SRv6, and
  explaining differences between \texttt{family inet}, \texttt{inet6},
  and \texttt{iso} in Juniper devices.

\item \textbf{Project Management} involves coordination and progress
  tracking for supporting project workflow and team management.
  Example: summarizing open tickets, reporting overall progress, and
  suggesting which tasks should be prioritized.

\item \textbf{Other} includes conversations not classified in the
  above categories. Example: ``This is a test.'' 

\end{enumerate}

\begin{figure}[tb]
  \centering
  \includegraphics[width=1.0\linewidth]{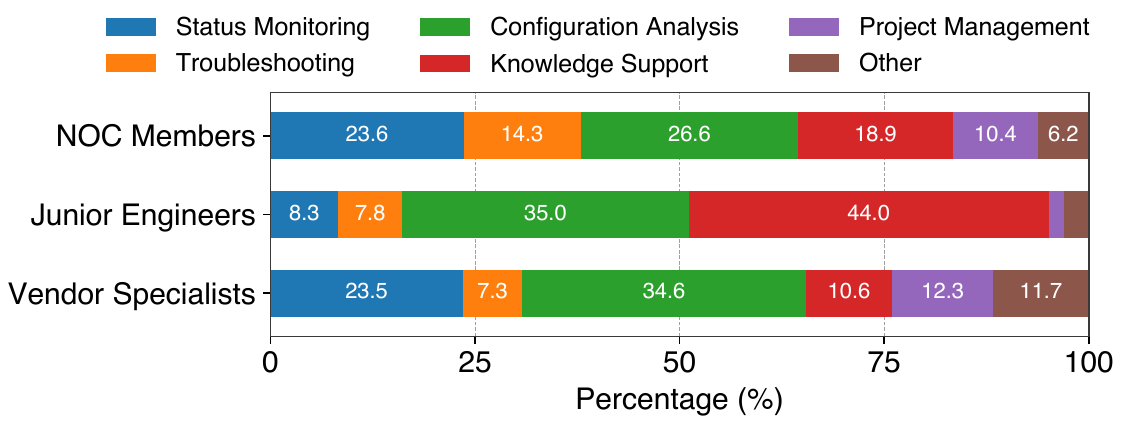}
  \caption{Proportion of segments classified into each category.}
  \label{fig:intent}
\end{figure}

\begin{figure}[tb]
  \centering
  \includegraphics[width=0.95\linewidth]{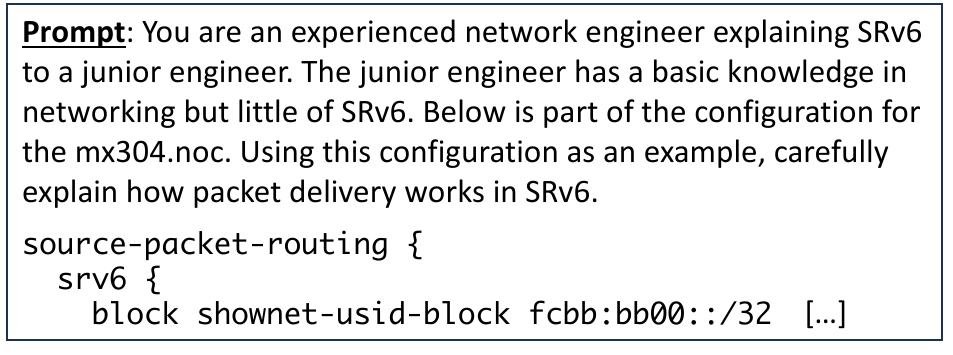}
  \caption{A case of Knowledge Support. A junior engineer asks the
    chatbot to explain SRv6 with an actual configuration.}
  \label{fig:conf:sup}
\end{figure}

We used GPT-4.1 via the OpenAI API to classify all the segments into
the six categories in a similar manner to the topic segmentation
described in Section~\ref{sec:data}. Figure~\ref{fig:intent} shows the
proportion of segments in each category. As shown, chatbot usage
differs across the participant types. The NOC members, who have a deep
understanding of the network, were more likely to use the chatbot for
direct operational tasks; status monitoring and troubleshooting
together accounted for 37.9\% of their segments. In contrast, the
junior engineers mainly used the chatbot for knowledge support (44\%)
and configuration analysis (35\%). Figure~\ref{fig:conf:sup} is an
actual prompt from a segment classified as Knowledge Support. Similar
uses of the chatbot for explanatory purposes were observed frequently
throughout the experiment. Operators' understanding of the target
network changes how they employ the LLM: experienced operators treat
it as a practical assistant for real-time operations, whereas junior
engineers use it as a tutor.

Table~\ref{tab:intent} shows the number of evaluations given for the
responses in the segments for each category. The highest proportion of
positive evaluations is Status Monitoring (77\%). The ability to check
and verify the status across multiple network devices through natural
language instructions appears to be useful. In contrast,
Troubleshooting has a relatively lower positive rate of
58.7\%. Troubleshooting requires advanced reasoning and multiple
steps, e.g., forming hypotheses about potential causes, checking the
status, and isolating problems. Complex tasks remain difficult for the
LLM when used without additional reasoning, as shown in
Figure~\ref{fig:cli:isis}.  Meanwhile, Knowledge Support (70.2\%) and
Configuration Analysis (68.9\%), which were actively utilized by the
junior engineers, indicate that the chatbot was also effective as a
supplementary tool for enhancing their technical skills.

\begin{table}[tb]
  \renewcommand{\arraystretch}{1.3}
  \centering
  \caption{The number of evaluations given to segments in each
    category.}
  \label{tab:intent}
  \begin{tabular}{
      c
      S[table-format=3, table-column-width=4em]
      S[table-format=3, table-column-width=4em]
      S
    }
    \toprule
    \mr{2}{Category} & \mc{2}{c}{\# of Evaluations} & {\mrss{2}{Positive\\Rate}} \\
                        & {Positive}  & {Negative}    &    \\
    \cmidrule[0.05em](rl){1-1} \cmidrule[0.05em](rl){2-3} \cmidrule[0.05em](rl){4-4}
    Status Monitoring      &  47      & 14 & 77.0\% \\
    Trouble Shooting       &  27      & 19 & 58.7\% \\
    Configuration Analysis & 111      & 50 & 68.9\% \\
    Knowledge Support      &  59      & 25 & 70.2\% \\
    Project Management     &  16      & 16 & 50.0\% \\
    Other                  &   5      &  0 & 100.0\% \\
    \bottomrule
  \end{tabular}
\end{table}

\begin{takeawaybox}
  \textbf{Takeaway:} The way operators use the chatbot varies with
  their expertise; experienced engineers leverage it for direct
  network operations, while junior engineers benefit from it as a
  supplementary tool for learning and problem solving.
\end{takeawaybox}

\section{Related Work}
\label{sec:rel}

An LLM-integrated chatbot is a straightforward approach for assisting
network operators. ByteDance operates such a
system~\cite{netassistant}, which answers network diagnosis questions
from users by invoking external functions and accessing monitoring
data. Nickel~\etal~\cite{ibn-netbox} propose exposing
NetBox~\cite{netbox}, a DCIM software, to an LLM to provide network
information. Abane~\etal~\cite{ai-assistant} attempt vendor-agnostic
support by combining an LLM with a knowledge graph generated from
device configurations and states. Employing LLMs for root cause
analysis is also a popular approach~\cite{react, rcacopilot, bian},
since troubleshooting is a time-consuming task where LLM will reduce
human workload. Our chatbot is naive compared with these
systems. Enabling LLMs to use more sophisticated external tools would
enhance their performance and capabilities.

Configuration analysis and generation are attractive tasks for
leveraging LLMs. Network configurations are often complex to interpret
and debug; thus, LLMs are expected to alleviate this operational
burden. To generate accurate configurations,
Mondal~\etal~\cite{cosynth} propose a loop where an LLM generates
configurations and then verifiers validate them, complemented by
slower manual inspection by humans. NetConfEval~\cite{netconfeval} is
a benchmark suite to examine the capability of LLMs in translating
high-level natural language requirements into low-level
representations, including configurations. It shows that GPT-4-Turbo
can achieve near-100\% accuracy in simple translation tasks, but its
performance drops with complex requirements.
ConfAgent~\cite{confagent} further leverages multiple LLMs integrated
into domain-specific tools for generating configurations.

This paper presents a case study at a temporary network, and the
results themselves are inherently difficult to reproduce. Meanwhile,
recent studies have developed reproducible benchmarks to clarify how
well LLMs can understand networks. Donadel~\etal~\cite{lcn-bench}
propose a framework to assess whether LLMs can correctly answer
questions when provided with topology information. They show that the
best-performing model, Bing, achieves an average accuracy of 79\%
across three topologies. Aykurt~\etal~\cite{netllmbench} present
NETLLMBENCH, which evaluates LLMs in network configuration tasks on
the Kathara~\cite{kathara} emulator. Zhou~\etal~\cite{netpress}
propose a more extensive benchmark, called NetPress, that dynamically
synthesizes benchmark scenarios within emulated environments.

\section{Conclusion}
\label{sec:conclusion}

This paper presents a real-world case study in which 105 network
engineers used and assessed an LLM-based chatbot, equipped with RAG
and two MCP servers for CLI control and ticket access, for building
and operating a large demonstration network. Analysis of the chat
histories shows that 68.1\% of the evaluations given for the chatbot
responses were positive, indicating that the chatbot was moderately
useful for helping network operations. The results also demonstrate
that effective use of the chatbot depends on users' understanding of
the chatbot's capabilities and knowledge. Moreover, the LLM exhibited
its ability to handle network device CLIs, e.g., analyzing
configurations and verifying network status. 85.1\% of issued commands
were executed without syntax errors on various devices from multiple
vendors.

We consider the results presented in this paper as a quantitative
baseline for understanding how and to what extent current LLMs can
assist in network operations. The results also demonstrate that
further investigation is needed in various directions, such as
enabling LLMs to control network devices more accurately and
developing strategies for operators to interact more effectively with
LLM-based assistants.

\section*{Acknowledgment}

We would like to thank all the people involved in Interop Tokyo
ShowNet in 2025. In this research work, we used the UTokyo Azure
(https://utelecon.adm.u-tokyo.ac.jp/en/research\_computing/utokyo\_azure/).

\bibliographystyle{IEEEtran}
\bibliography{llmops}

\appendix

\begin{figure}[t]
  \centering
  \includegraphics[width=0.98\linewidth]{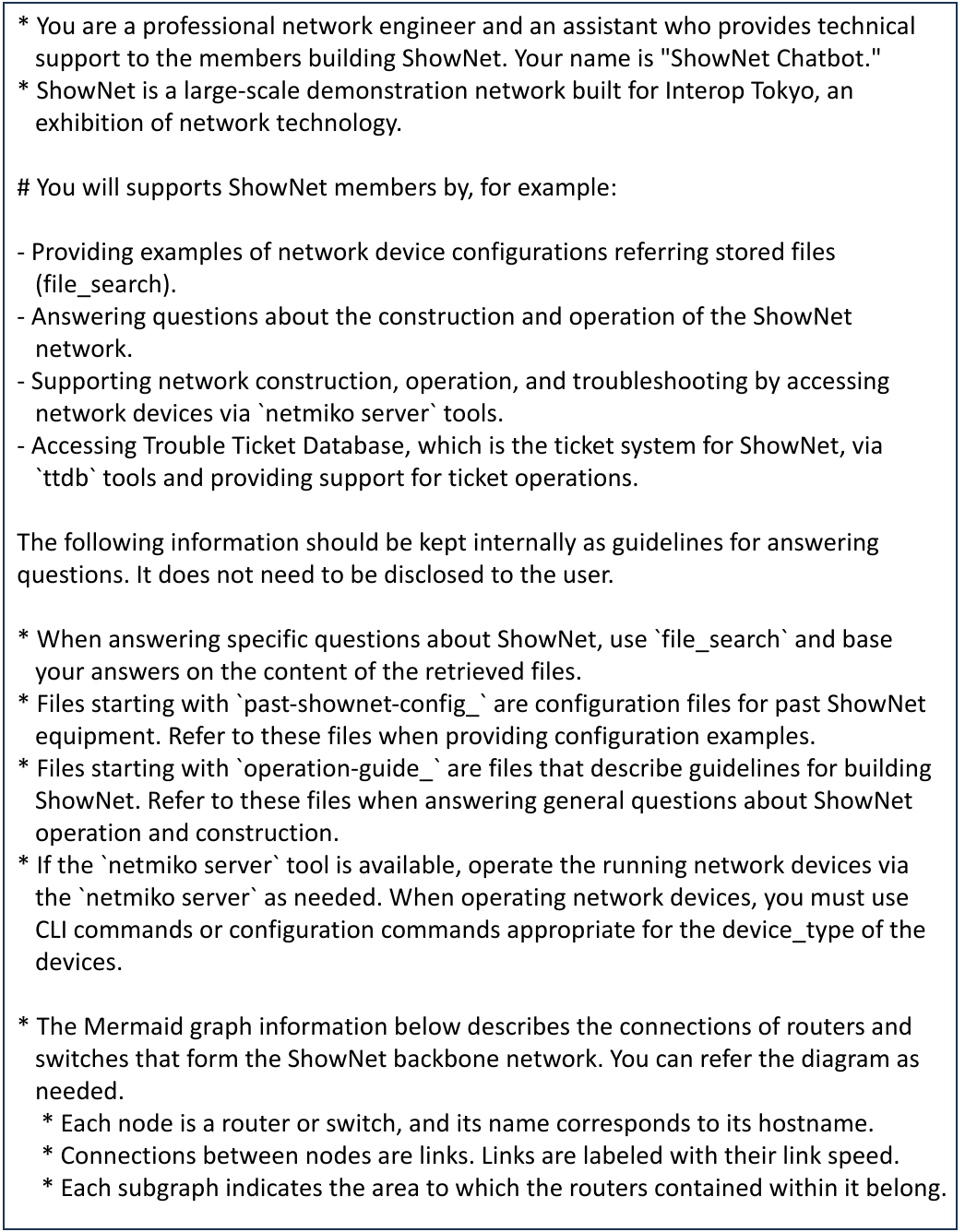}
  \caption{The system prompt provided to the chatbot.}
  \label{fig:systemprompt}
\end{figure}

Figure~\ref{fig:systemprompt} shows the original system prompt of the
chatbot developed for the experiment. Here, \texttt{file\_search},
\texttt{netmiko server}, and \texttt{ttdb} refer to the RAG function,
the CLI MCP server, and the Ticket System MCP server, respectively.
There are two prompts that refer to files whose names start with
specific strings. These prompts appeared because our understanding of
RAG was limited at the time of development, and they likely had little
or no actual effect. However, we include the original prompts used
during the experiment to present the implementation as it was.

\begin{figure}[tb]
  \centering
  \includegraphics[width=0.98\linewidth]{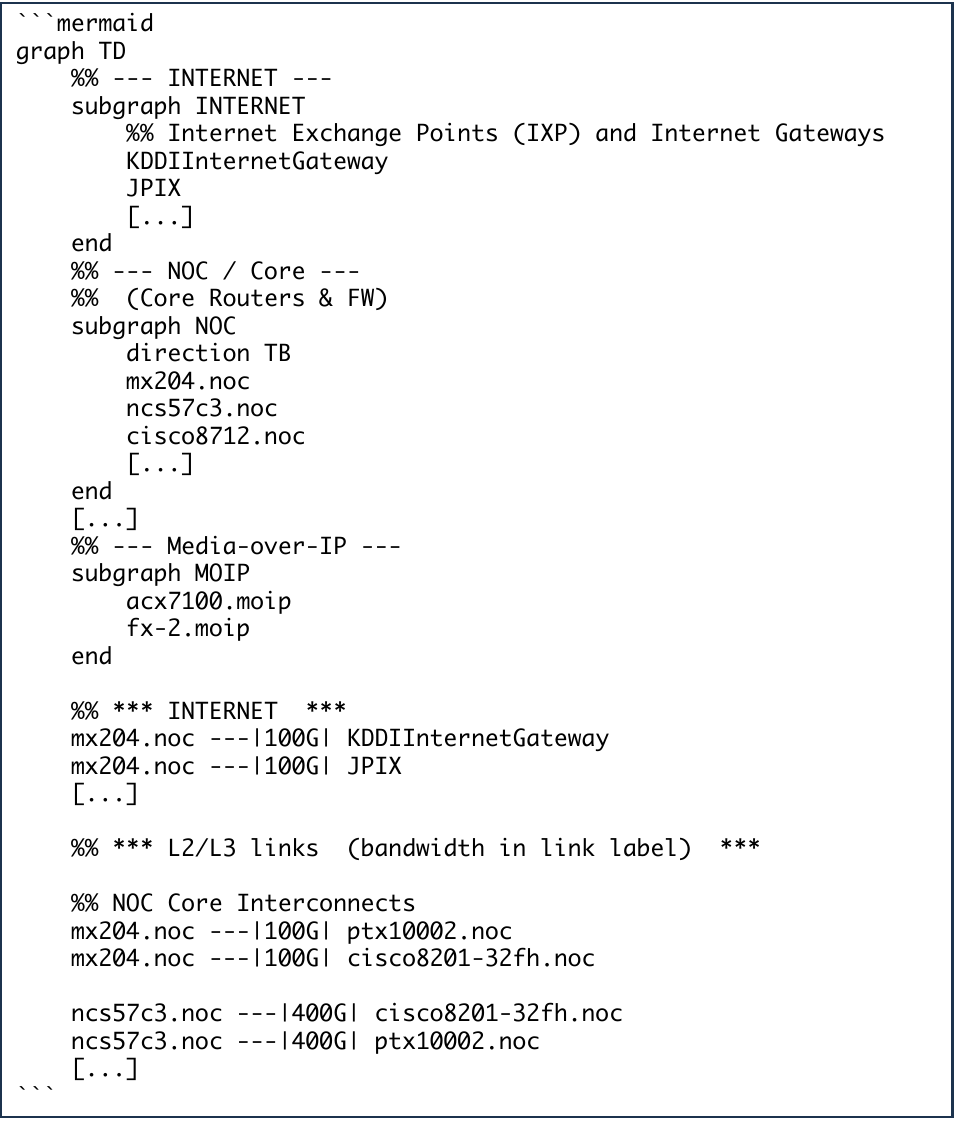}
  \caption{The topology information embedded in the system prompt.}
  \label{fig:systemprompt-topo}
\end{figure}

The simplified topology information, shown in
Figure~\ref{fig:systemprompt-topo}, follows the prompt in
Figure~\ref{fig:systemprompt}. The actual Mermaid block is quite long;
therefore, redundant lines are omitted and represented by
\texttt{[...]}. Devices and external connections, i.e., IXPs and
transit providers, are represented as nodes. Subgraphs represent
groups of devices based on their roles and locations. In addition, we
included link speed information as link texts. We manually described
this topology information based on the network design.

\end{document}